\documentclass[11pt,a4paper]{article}
\pdfoutput=1 
\usepackage{jheppub} 
\usepackage[T1]{fontenc} 
\usepackage{amsmath} \usepackage{amssymb} \usepackage{amsfonts}
\usepackage{amsthm} \usepackage{mathtools} \usepackage{mathrsfs}
\usepackage{nicefrac}
\usepackage{graphicx}
\usepackage{paralist}
\usepackage[caption=false, singlelinecheck=false, justification=raggedright]{subfig}%
\usepackage{xspace} \usepackage{txfonts} \usepackage{pifont}
\usepackage{multirow} \usepackage{array} \usepackage[normalem]{ulem}%
\usepackage{rotating} 
\usepackage{lscape}
\usepackage{lipsum}

\begin{document}

\title{Probing new physics scenarios of muon \texorpdfstring{$g-2$}{g-2} via
\texorpdfstring{$J/\psi$}{J/psi} decay at BESIII}

\author[a]{Gorazd~Cveti\v{c},}%
\author[b,c]{C.S.~Kim,}%
\author[c]{Donghun Lee,}%
\author[c]{and Dibyakrupa Sahoo}%

\affiliation[a]{Department of Physics, Universidad Tecnica, Federico Santa
Maria, Valpraiso, Chile}%
\affiliation[b]{Institute of High Energy Physics, Dongshin University, Naju
58245, Korea}%
\affiliation[c]{Department of Physics and IPAP, Yonsei University, Seoul 03722,
Korea}%

\emailAdd{gorazd.cvetic@usm.cl}%
\emailAdd{cskim@yonsei.ac.kr}%
\emailAdd{donghun.lee@yonsei.ac.kr}%
\emailAdd{sahoodibya@yonsei.ac.kr}%

\date{\today}

\abstract{The disagreement between the standard model prediction and the
experimental measurement of muon anomalous magnetic moment can be alleviated by
invoking an additional particle which is either a vector boson ($X_1$) or a
scalar ($X_0$). This new particle, with the  mass $m_X \lesssim 2 m_\mu$, can be
searched for in the decay $J/\psi \to \mu^- \mu^+ X$, where $X$ is missing. Our
numerical study shows that the search is quite feasible at the BESIII experiment
in the parameter space allowed by muon $g-2$ measurements.}




\maketitle

\section{Introduction}
\label{sec:intr}

The magnetic moment of muon ($\boldsymbol\mu$) is directly proportional to its
intrinsic spin ($\mathbf{S}$), $\boldsymbol\mu = g_\mu (\nicefrac{e}{2m_\mu} )
\mathbf{S}$, where $g_\mu, e, m_\mu$ are the $g$-factor, elementary charge and
mass of muon, respectively. The Dirac equation predicts $g_\mu=2$ since muon is
an elementary (i.e.\ structureless) spin-$\nicefrac{1}{2}$ fermion. However,
radiative corrections from quantum loops are known to result in a tiny but
non-zero deviation from this value. This deviation is quantified by the
anomalous magnetic moment, $a_\mu = \left(g_\mu - 2\right)/2$. The anomalous
magnetic moment of muon has been very precisely measured by the E821 experiment
at Brookhaven National Laboratory (BNL)~\cite{Bennett:2006fi}. The experimental
measurement is found to be about $3.3\sigma$ larger than the Standard Model (SM)
prediction \cite{Tanabashi:2018oca},
\begin{equation}\label{eq:DelA}
\Delta a_\mu \equiv a_\mu^{\textrm{exp}} - a_\mu^{\textrm{SM}} = \left(261 \pm 63
\pm 48 \right) \times 10^{-11},
\end{equation}
where the first error is from experiment and the second one is from theory
prediction. This result, as well as the recent observation that inclusion of SM
radiative corrections is not sufficient to resolve the anomaly in $a_\mu$
\cite{Campanario:2019mjh}, can be considered as possible hints of some
underlying new physics. Further, the contribution of the leading order hadronic
vacuum polarization to the muon $a_{\mu}$, extracted with high accuracy from the
measurements of $e^+ e^- \to$ hadrons \cite{Jegerlehner:2017lbd}, also cannot
reduce the present discrepancy $\Delta a_{\mu}$. In this paper, we shall probe
two simple new physics scenarios, involving either a new vector boson (say
$X_1$) or a new scalar (say $X_0$), which can contribute to muon anomalous
magnetic moment and alleviate the existing discrepancy between theory prediction
and experimental measurement. Our proposed search for the new particle is via
the study of the decay $J/\psi \rightarrow \mu^- \mu^+ X_{0,1}$, in the range
$m_X \leqslant 2m_\mu$, with $X_{0,1}$ as the invisible or missing final state.
This decay mode can be studied in the ongoing BESIII experiment or in any future
experiment which can produce large number of on-shell $J/\psi$ mesons at rest.

The experiment BESIII has clear advantages for the studies of production of such
$X$ particles in comparison with the experiments Belle II and BaBar where the
continuum $e^+ e^- \to \mu^+ \mu^- X$ process can be considered \cite{BelleII}.
First, at BESIII a very large number ($\sim 10^{11}$) of on-shell $J/\psi$
particles (at rest) will be produced \cite{BESIII,Yuan:2019zfo,Ablikim:2019hff},
and consequently the number of $J/\psi \to \mu^+ \mu^- X$ events that can be
produced is significantly higher than that at Belle II or BaBar. This is due to
the fact that the continuum process $e^+ e^- \to \mu^+ \mu^- X$ at Belle II or
BaBar does not take place via on-shell exchanged particle, making the number of
such events considerably suppressed. Secondly, the center-of-mass energy
$\sqrt{s}$ of the events at Belle II and BaBar is very high ($\approx 10$~GeV),
making the cross section of the considered process suppressed, $\sigma \propto
1/s$. In addition, since BESIII has the final state kinematics strongly
constrained by the on-shell $J/\psi$ (which has a very small decay width)
without any initial soft photon radiation $\gamma_{_{\rm ISR}}$, the background
effects are much easier to analyze than at Belle II or BaBar.

We would like to point out that our approach to study the bosonic mediators
$X_{0,1}$ differs from another recent proposal~\cite{Jiang:2018jqp}, which
studied $J/\psi$ decay at BESIII through the process of $J/\psi \rightarrow
X_{0,1} + \gamma \rightarrow \mu^- \mu^+ \gamma$ in the range $m_{J/\psi}
\geqslant m_X \geqslant 2m_\mu$:
\begin{inparaenum}[(1)]
\item Unlike our paper, Ref.~\cite{Jiang:2018jqp} does not consider the muon
anomalous magnetic moment to probe and constrain the parameter space for
$X_{0,1}$. %
\item In Ref.~\cite{Jiang:2018jqp} the mediators $X_{0,1}$ can be scalar,
pseudo-scalar, vector or axial-vector; while in our case, in order to explain
the observed muon anomalous magnetic moment, we have constrained ourselves to
the scalar and vector possibilities only, thus making the scenario much simpler.
\end{inparaenum}
Probing such light scalar and vector particles have also been discussed in
context of other decay modes in
Refs.~\cite{Correia:2016xcs,Correia:2019pnn,Correia:2019woz} in context of
specific $U(1)$ extensions of the SM.

Our paper is organized as follows. In Sec.~\ref{sec:theory} we elaborate on the
two new physics scenarios under our consideration and clearly lay down the
search strategy using the decay $J/\psi \rightarrow \mu^- \mu^+ X_{0,1}$. This
is followed in Sec.~\ref{sec:numerical} by a numerical study of the two
scenarios as well as that of the competing SM background processes, in context
of the BESIII experiment. Finally we conclude in Sec.~\ref{sec:conclusion}
emphasizing the various salient features of our study.

\section{Theoretical motivation and experimental search strategy}\label{sec:theory}
\label{sec:tmss}

\subsection{Simplest muonic interactions}
\label{subs:smi}

For an effective solution to the problem of muon anomalous magnetic moment
without affecting any other existing studies, it would be ideal if the new
interactions that get introduced only involve muons. In this context it is well
known that if there exists either a scalar $X_0$ or a vector $X_1$ that
interacts only with the muons, we can write down the following interaction
Lagrangians:
\begin{subequations}\label{eq:int-lagrangian}
\begin{align}
\mathcal{L}_{\mu}^{\textrm{scalar}} &= - g_0 \, X_0 \; \overline{\mu}\,\mu,
\label{eq:int-lagrangian-s}\\%
\mathcal{L}_{\mu}^{\textrm{vector}} &= - g_1 \, X_{1\alpha} \;
\overline{\mu}\,\gamma^\alpha\,\mu. \label{eq:int-lagrangian-v}%
\end{align}
\end{subequations}
\begin{figure}[tbph]
\centering%
\includegraphics[scale=1]{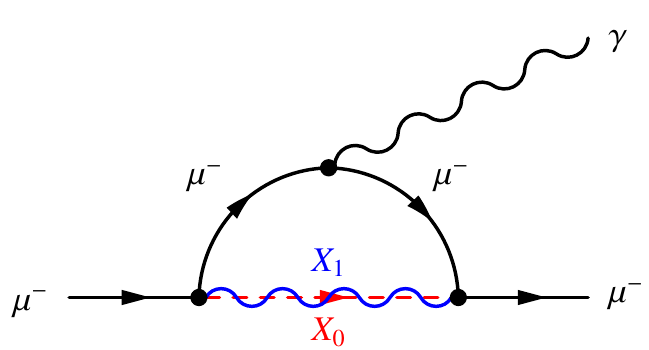}
\vspace{-5mm}
\caption{Contribution from $X_0$ or $X_1$ to muon anomalous magnetic moment.}
\label{figamu}
\end{figure}
These interactions give contributions to the muon anomalous magnetic moment. The
leading order contributions, from the loops as shown in Fig.~\ref{figamu}, are
given by~\cite{Review},
\begin{subequations}\label{eq:DelA-sv}
\begin{align}
\Delta a_\mu^{\textrm{scalar}} &= \frac{g_0^2}{8\pi^2} \int_{0}^{1} dx \,
\frac{m_\mu^2 (1-x)(1-x^2)}{ m_\mu^2(1-x)^2+m_X^2 x},\label{eq:DelA-s}\\%
\Delta a_\mu^{\textrm{vector}} &= \frac{g_1^2}{8\pi^2} \int_{0}^{1} dx \,
\frac{2m_\mu^2 \, x \, (1-x)^2}{m_\mu^2 (1-x)^2 + m_X^2 x},\label{eq:DelA-v}
\end{align}
\end{subequations}
where $m_X$ is used to denote the mass of both $X_0$ and $X_1$ and these results
are applicable for $m_X \lesssim 2 m_\mu$. The region of parameter space in
$g_{0,1}$-$m_X$ planes allowed by the current discrepancy in anomalous magnetic
moment (at $2\sigma$ level, adding the errors in Eq.~\eqref{eq:DelA} by
quadrature) is shown in Fig.~\ref{fig:g01mX}. It must be noted that the
condition $m_X < 2m_\mu$ is imposed to kinematically forbid the only possible
tree-level decay $X_{0,1} \to \mu^- \mu^+$. Other decay modes, such as $X_{0,1}
\to e^- e^+, \nu_\ell \overline{\nu}_\ell$ for $\ell=e,\mu,\tau$ are not allowed
at the tree level, but in principle these are possible via loop processes which
are suppressed if not forbidden kinematically.

\begin{figure}[hbtp]
\centering%
\includegraphics[scale=0.8]{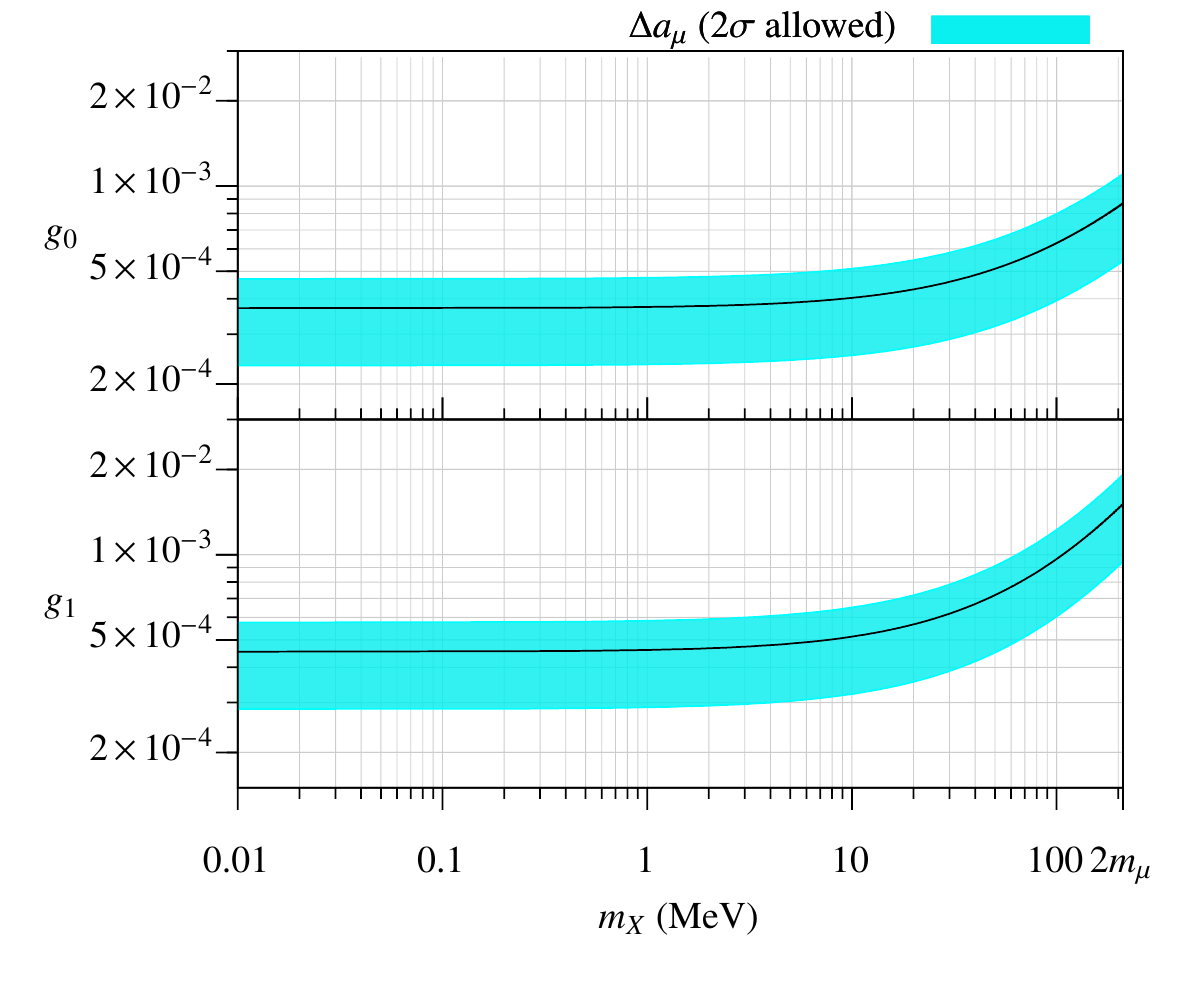}%
\vspace{-5mm}%
\caption{The parameters $g_0$, $g_1$ and $m_X \leqslant 2m_\mu$, as allowed by
$\Delta a _\mu$ at $2\sigma$ level. Obtained by adding the errors in
Eq.~\eqref{eq:DelA} in quadrature and by using Eq.~\eqref{eq:DelA-sv}. It is
important to note here that $X_{0,1}$ are assumed to exclusively interact with
muons alone as per Eq.~\eqref{eq:int-lagrangian}.}%
\label{fig:g01mX}
\end{figure}

It is important to note that we have considered in Eq.~\eqref{eq:int-lagrangian}
only parity even scenarios here, i.e.\ no pseudo-scalar or axial-vector
possibilities are being considered. This is so because of the fact that the
contribution of pseudo-scalar and axial-vector particles to the anomalous
magnetic moment of muon has opposite sign which makes the discrepancy between
theory prediction and experimental measurement much larger. Therefore, we shall
limit ourselves to scalar and vector cases only. %
It is also interesting to note that the scalar and vector scenarios may have
their origin in a more complete model. There are many UV complete models which
can accommodate Eqs.~\eqref{eq:int-lagrangian}, but they have additional
features which might constrain the model severely. In the next subsection we
shall consider one of the simplest probable models elucidating the main ideas
behind such an approach.

\subsection{Towards a possibly complete model: \texorpdfstring{$U(1)_{L_{\mu} - L_{\tau}}$}{U(1) Lmu-Ltau}}
\label{subs:cm}

Here we are concerned with a specific extension of the SM gauge group, namely by
an additional symmetry group $U(1)_{L_\mu-L_\tau}$ which conserves the
difference between the muon and tau lepton numbers, while keeping the overall
model anomaly free~\cite{He:1990pn,He:1991qd,Foot:1994vd}. The new gauge
symmetry, under which only the second and third generations of leptons are
charged, gives rise  to an additional massive vector gauge boson, $X_1$, which
naturally couples to second and third generations of leptons alone at the tree
level. The mass of the $X_1$ boson, $m_X$, can be generated via either
spontaneous symmetry breaking or Stueckelberg
mechanism~\cite{Ruegg:2003ps,Feldman:2006wb}. The underlying Lagrangian
including the kinetic term, mass term and gauge interaction term, for the gauge
boson $X_1$ is therefore given by,
\begin{equation}\label{eq:lagr}
\mathcal{L} \supset \mathcal{L}_{\textrm{SM}} - \frac{1}{4} X_1^{\alpha\beta}
X_{1\alpha\beta} + \frac{m_X^2}{2} X_1^\alpha X_{1\alpha} - X_{1\alpha}
J^\alpha_{\mu-\tau},
\end{equation}
where $X_{1\alpha\beta} \equiv \partial_\alpha X_{1\beta} - \partial_\beta
X_{1\alpha}$ is the field strength tensor, and $J^\alpha_{\mu-\tau}$ is the $\mu - \tau$
current given by,
\begin{equation}\label{eq:current-X}
J^\alpha_{\mu-\tau} = g_1 \left( \overline{\mu} \gamma^\alpha \mu - \overline{\tau}
\gamma^\alpha \tau + \overline{\nu}_{\mu} \gamma^\alpha P_L \nu_{\mu} -
\overline{\nu}_{\tau} \gamma^\alpha P_L \nu_{\tau} \right),
\end{equation}
where $P_L \equiv \tfrac{1}{2}\left(1-\gamma^5\right)$ is the left projection
operator. The first term in $J^\alpha_{\mu-\tau}$ in Eq.~\eqref{eq:current-X} is
same as the term in Eq.~\eqref{eq:int-lagrangian-v}. However, in the
$U(1)_{L_\mu - L_\tau}$ extension of the Standard Model, there appear additional
terms which would contribute to the anomalous magnetic moment of tau via a loop
diagram similar to the one in Fig.~\ref{figamu}. However, currently the
anomalous magnetic moment of tau is not well measured to constrain these new
physics scenarios~\cite{Tanabashi:2018oca}. Therefore, we shall refrain from
using anomalous magnetic moment of tau in this paper. Nevertheless, in the
$U(1)_{L_\mu-L_\tau}$ extension, the $X_1$ vector boson is not necessarily
stable even for $m_X < 2 m_\mu$ as $X_1 \to \overline{\nu}_\mu \nu_\mu,
\overline{\nu}_\tau \nu_\tau$ are allowed at the tree level. The decay $X_1 \to
e^- e^+$ is still forbidden at the tree level and can only happen (if
kinematically allowed) via quantum loop and it would therefore be suppressed.
Thus, predominantly the $X_1$ boson would decay invisibly and its direct
signature at experiments would be missing 4-momentum. Further, $X_1$ could also
couple to dark matter constituting yet another invisible decay mode.

\begin{figure}[hbtp]
\centering%
\includegraphics[scale=0.8]{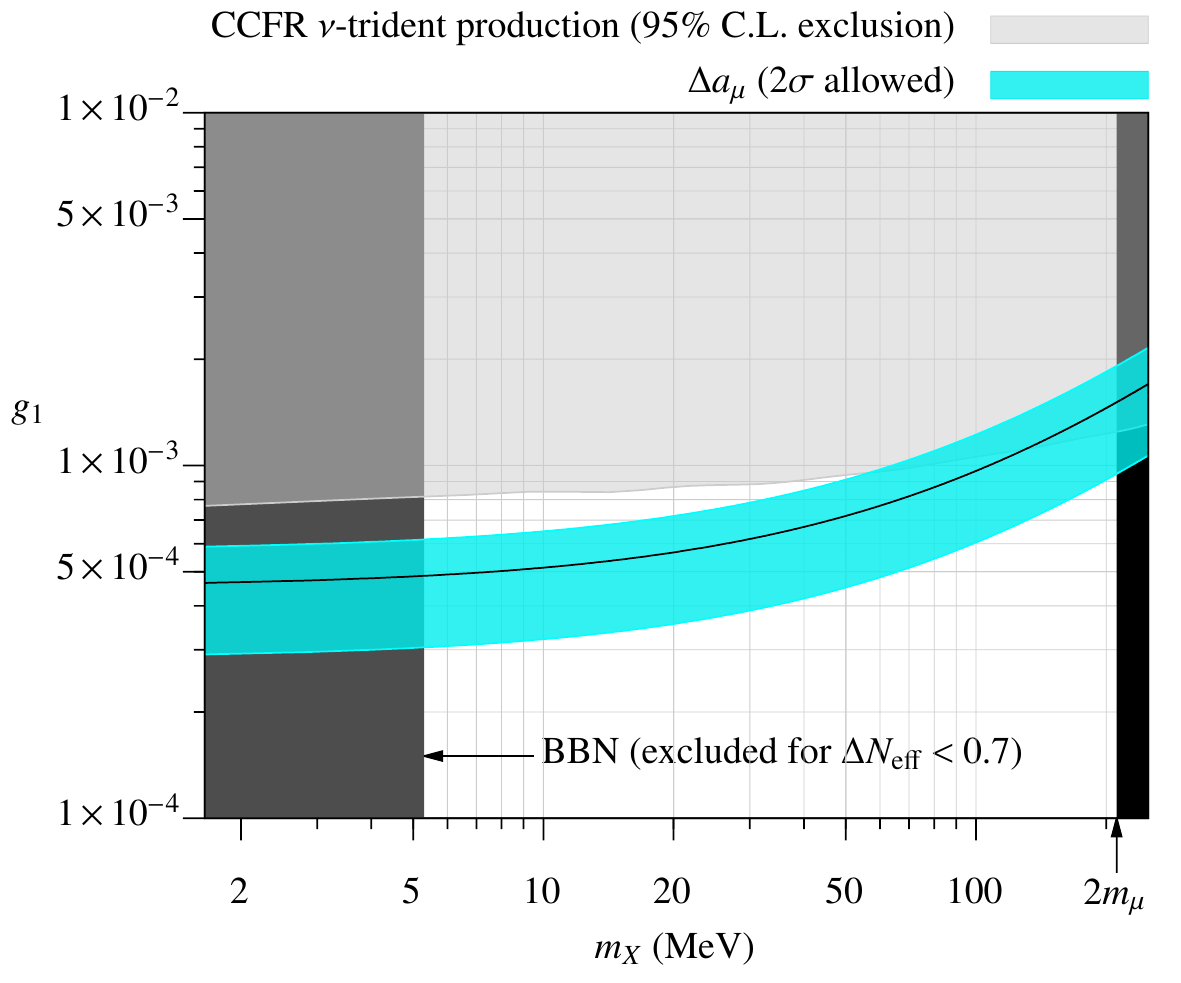}%
\vspace{-4mm}%
\caption{Allowed region in the $g_1$-$m_X$ plane from anomalous magnetic moment
of muon. Here we have combined the experimental and theoretical uncertainties in
quadrature.}%
\label{fig:gXmX}
\end{figure}

Considering only the $U(1)_{L_\mu-L_\tau}$ new physics and depending on whether
the mass $m_X$ is small or large, there exist other diverse observations which
can also probe or constrain the allowed region in the $g_1$-$m_X$ plane, such
as
\begin{inparaenum}[(1)]
\item precision measurements at $Z$ pole \cite{BelleII}, \item neutrino-nucleus
scattering involved in neutrino trident
production~\cite{Mishra:1991bv,Altmannshofer:2014pba}, \item rare kaon decays in
beam-dump experiments \cite{kaonc}, \item tests of lepton universality such as
$R(K)$, $R(K^*)$~\cite{Gauld:2013qja,DAmico:2017mtc,DAmbrosio:2019tph} etc., as
well as \item big-bang nucleosynthesis with constraint on deviation from
effective number of light neutrinos ($\Delta N_\textrm{eff}$) \cite{BBN1,BBN2}.
\end{inparaenum}
In the mass range of our interest, i.e.\ $m_X < 2m_\mu$, the constraints from
neutrino trident production as measured by the CCFR
experiment~\cite{Mishra:1991bv} as well as the constraint from big-bang
nucleosynthesis~\cite{BBN1} are relevant and are shown in
Fig.~\ref{fig:gXmX}. It is very clear from Fig.~\ref{fig:gXmX} that the
exclusion region from neutrino trident production experiment strongly
constrains a portion of the parameter space allowed by muon anomalous
magnetic moment in the higher mass ranges.

In order to probe the allowed parameter space in a more thorough manner, our
chosen process must not only have high yield, but should have a distinct
experimental signature in the parameter region of interest. In the next
subsection, we analyze a decay mode which satisfies these criteria.

\subsection{New search strategy}
\label{subs:nss}

Since we are concerned with probing the scalar $X_0$ and vector boson $X_1$
(which can be called ``muon-philic'') as they satisfy the
Eq.~\eqref{eq:int-lagrangian}, it is only natural to think of a process that
involves muons in the final state to search for $X_{0,1}$. Moreover, as we have
discussed above, $X_{0,1}$ with mass $m_X < 2 m_\mu$ would be fully invisible as
it is electrically neutral and stable (if it decays, then it decays to
neutrino-antineutrino pair and possibly to dark matter
particles, which are also invisible). Thus, the process we consider must have
missing 4-momentum in the final state, and it should be possible, in principle
and practice, to measure the missing 4-momentum as precisely as experimentally
possible. An excellent process that satisfies all these criteria is the decay
$J/\psi \rightarrow\mu^- \mu^+ X_{0,1}$, where $J/\psi$ needs to be produced at
rest so that the initial 4-momentum is fully known and fixed. Because the final
state has two muons which are well reconstructed in modern detectors, this would
imply that the missing 4-momentum can be precisely inferred in such a case.

It is important to note that
\begin{inparaenum}[(1)]
\item extremely large sample of on-shell $J/\psi$ can be produced in $e^+ e^-$
colliders such as BESIII, which provides statistically significant number of
signal events, %
\item the extremely narrow width of $J/\psi$ ensures that events with the
missing initial soft photon radiation $\gamma{_{\rm ISR}}$ from the colliding
electron-positron beams can be safely ignored (unlike the continuum process of
$e^- e^+ \rightarrow \mu^- \mu^+ X_{0,1}$, where the initial state radiation of
soft photons would be a major background) and %
\item the missing final state soft photon radiation from the muons (as shown in
Fig. 4(c)), which constitutes the dominant background for our decay $J/\psi
\rightarrow \mu^- \mu^+ X_{0,1}$, can also be dealt with very precisely due to
the fact that $J/\psi \to \mu^- \mu^+$ is very well studied and the missing mass
in $J/\psi \to \mu^- \mu^+ \gamma_\textrm{soft}$ events always peaks at the
photon pole, i.e.\ at missing mass equal to zero.
\end{inparaenum}
The continuum process $e^- e^+ \rightarrow \mu^- \mu^+ X_{0,1}$, which can be
studied at experiments such as Belle~II has a much larger set of background
processes and the strategy to be dealt with such a study can be found in
Ref.~\cite{BelleII}.

\begin{figure}[tb]
\centering%
\includegraphics[width=\linewidth,keepaspectratio]{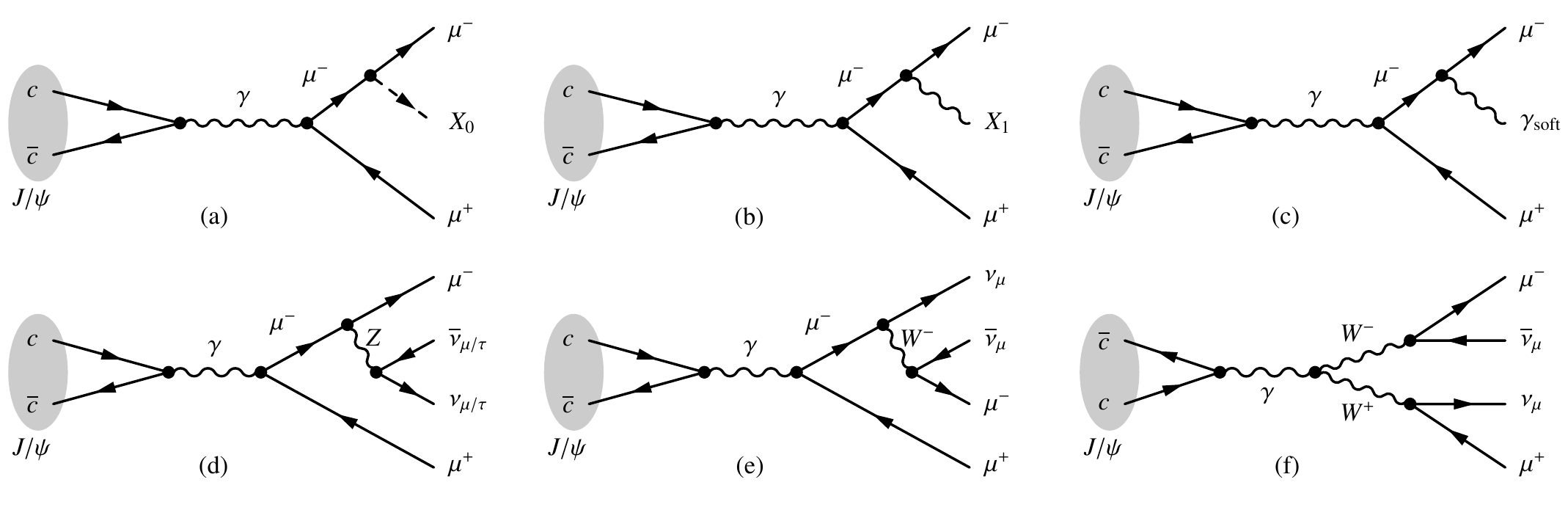}%
\vspace{-5mm}%
\caption{Quark-level diagrams for the new physics signal of $X_{0}$ or $X_{1}$
(a,b) and the dominant background from final state radiation of soft-photon (c)
as well as the sub-dominant background processes (d,e,f) in the Standard Model.
Diagrams (a), (b), (c), (d) and (e) have conjugate diagrams (with vertices on
the $\mu^+$ line) which are not drawn here. Charge conjugation invariance
prevents radiation of soft photon from $J/\psi$.}%
\label{fig:feynman-diagrams}
\end{figure}

The quark-level Feynman diagrams for the signal and background processes are
shown in Fig.~\ref{fig:feynman-diagrams}. It should be noted that, except the
final state soft radiation (shown in Fig.~\ref{fig:feynman-diagrams}(c)), the
other backgrounds (shown in Fig.~\ref{fig:feynman-diagrams}(d,e,f)) are
extremely suppressed as they involve two or more weak vertices. Quantitatively,
the soft photon background dominates over the weak background by roughly eight
orders of magnitude. Therefore, we shall not dwell upon any of the weak
background processes, shown in Fig.~\ref{fig:feynman-diagrams}, in our numerical
studies.

\begin{figure}[ht!]
\centering%
\includegraphics[width=0.9\linewidth,keepaspectratio]{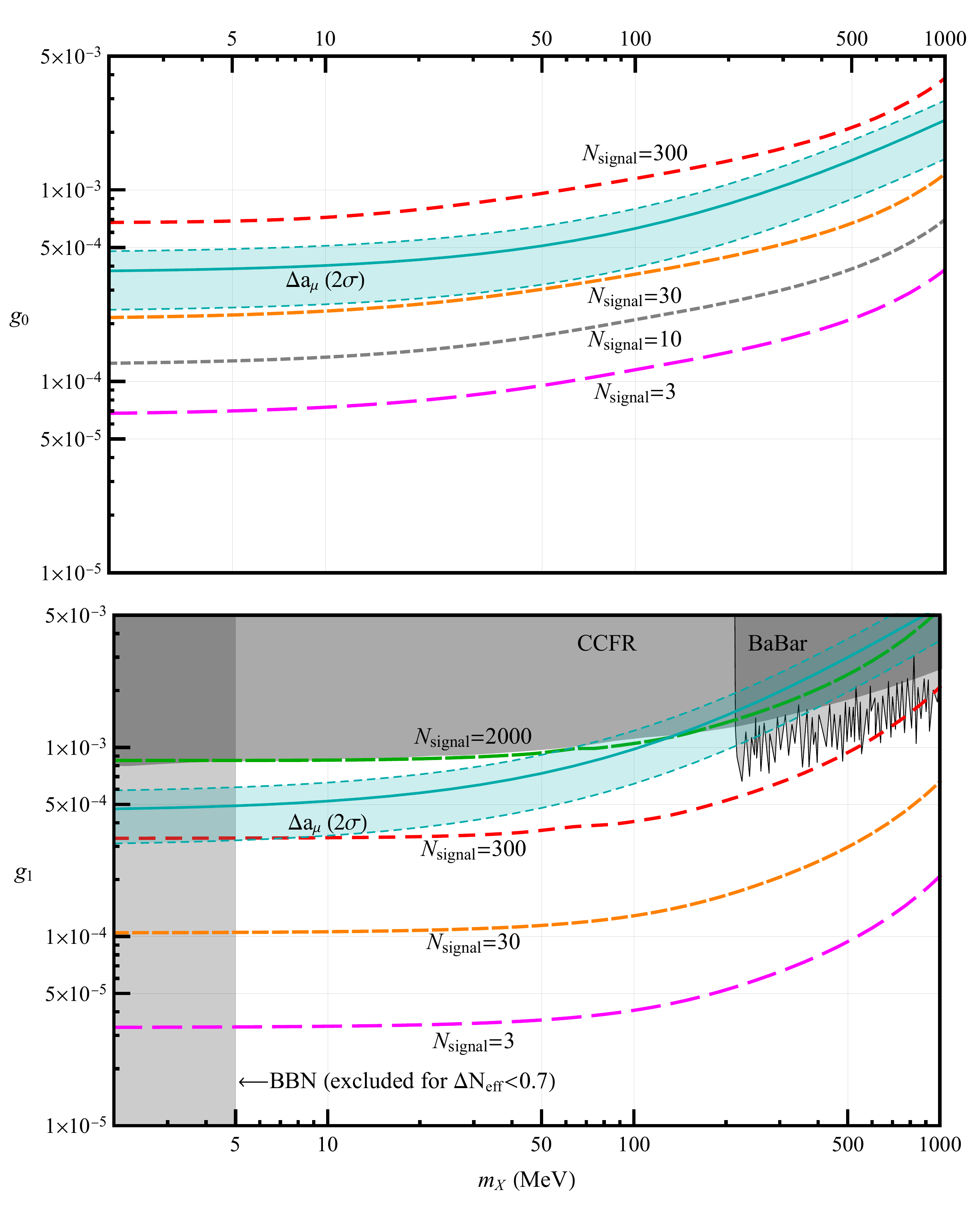}%
\vspace{-5mm}%
\caption{Estimate of number of $J/\psi \to \mu^- \mu^+ X_{0,1}$ events in the
$g_{0,1}$-$m_X$ parameter space in context of BESIII experiment. The parameter
space allowed by $\Delta a _\mu$ at $2\sigma$ level would give rise to about
$300$-$2000$ signal events for $J/\psi \to \mu^- \mu^+ X_1$ decay, and
$30$-$300$ signal events for $J/\psi \to \mu^- \mu^+ X_0$ decay at the BESIII
experiment. These numbers are obtained after considering an energy cut, which
will be discussed in Sec.~\ref{subs:me}.}%
\label{fig:event-estimate}
\end{figure}

Now, considering the signal events alone in the context of the BESIII experiment
where it is estimated that $10^{11}$ number of $J/\psi$ would be produced, we
find that roughly $300$-$2000$ events of $J/\psi \to \mu^- \mu^+ X_1$ or about
$30$-$300$ events of $J/\psi \to \mu^- \mu^+ X_0$ can be expected (see
Fig.~\ref{fig:event-estimate}), corresponding to the region of parameter space
allowed by the muon anomalous magnetic moment. Due to this large number of
events expected, if we observe fewer or no events, then this would rule out the
simplistic scalar and vector explanation of muon anomalous magnetic moment which
we considered here. In order to fully understand the feasibility of the decay
$J/\psi \to \mu^- \mu^+ X_{0,1}$, we made a comparative study of the dominant SM
background and the new physics events in the following Section.

\section{Numerical study of feasibility of probing the new physics scenarios}\label{sec:numerical}

As we have mentioned in the previous Section, the final state radiation of soft
photon from either of the muon lines is the dominant as well as the only
relevant background in our case. We have devised a non-traditional approach to
study the dominant background process $J/\psi \to \mu^- \mu^+
\gamma_\textrm{soft}$. Since the soft photon is not detected, the observed
events are essentially $J/\psi \rightarrow \mu^- \mu^+$ events as the
soft-photon has low energy (BESIII cannot detect soft photons with energy $<
20$~MeV~\cite{Ablikim:2009aa}) and hence it also has a small magnitude of
3-momentum, essentially keeping the two final muons back-to-back, within the
accuracy of the experimental resolution of the muon tracks (provided $J/\psi$ is
produced at rest, which is true for the BESIII experiment). Therefore, if we
take the muon momentum resolution of the experiment into account for the muon
pair in $J/\psi \rightarrow \mu^- \mu^+$, we can essentially get all possible
soft-photon events as required for our numerical study. At BESIII the error in
the momentum resolution is about $1\%$ of the momentum being
measured~\cite{Ablikim:2009aa}, i.e.\ the experimental uncertainty in the
measurement $\sigma_p \sim 0.01 p$ where $p$ is the central value of the
magnitude of the 4-momentum. For simplicity, we have specifically assumed
$\sigma_p \sim 15~\rm MeV$ for our numerical simulation, which would presumably
provide bigger momentum uncertainty for most events than what is expected
experimentally. We utilize a multitude of observables and relevant cuts, as
discussed below, to distinguish the signal and background events so as to
facilitate the discovery of new physics in our decay mode $J/\psi \rightarrow
\mu^- \mu^+ X_{0,1}$. For our numerical study we have considered $10^{11}$
number of $J/\psi$ that would be produced at rest at the BESIII experiment. In
the numerical simulation of signal events we have also considered the central
value of $g_{0,1}$ for the corresponding value of $m_X$ as allowed by the muon
anomalous magnetic moment, see Fig. 5. Below we first illustrate our methodology
for the vector boson case and finally discuss the scalar case and compare.

\subsection{Probing the vector boson case}
\label{subs:pvb}

\subsubsection{Square of the missing mass}
\label{ssub:smm}

If we denote the 4-momenta of $J/\psi,~ \mu^-,~\mu^+$ and the missing component
(which can be the new particles $X_{0,1}$ or soft photon $\gamma_\textrm{soft}$) by
$p_J,~ p_-,~ p_+$ and $p_\textrm{miss}$, then the missing
mass $m_\textrm{miss}$ is given by,
\begin{equation} \label{eq:mmiss}
m_\textrm{miss}^2 \equiv p_\textrm{miss}^2 = \left(p_J - p_- - p_+ \right)^2.
\end{equation}
Theoretically $m_\textrm{miss}^2$ distribution for signal events will be
characterized by very sharp peaks due to the tiny decay width of $X_{0,1}$,
e.g.\ in the $U(1)_{L_\mu - L_\tau}$ model the $X_1 \rightarrow \nu_{\mu}
\overline{\nu}_{\mu}, \nu_{\tau} \overline{\nu}_{\tau}$ decays provide
$\Gamma_{X_1} = g_1^2 \, m_X/(12\pi)$, and the background events should be
crowded at $m_\textrm{miss}^2=0$. However errors in measurements of momenta will
smear the distribution of events for both signal and background.

\begin{figure}[htb]
\centering%
\includegraphics[width=0.8\linewidth]{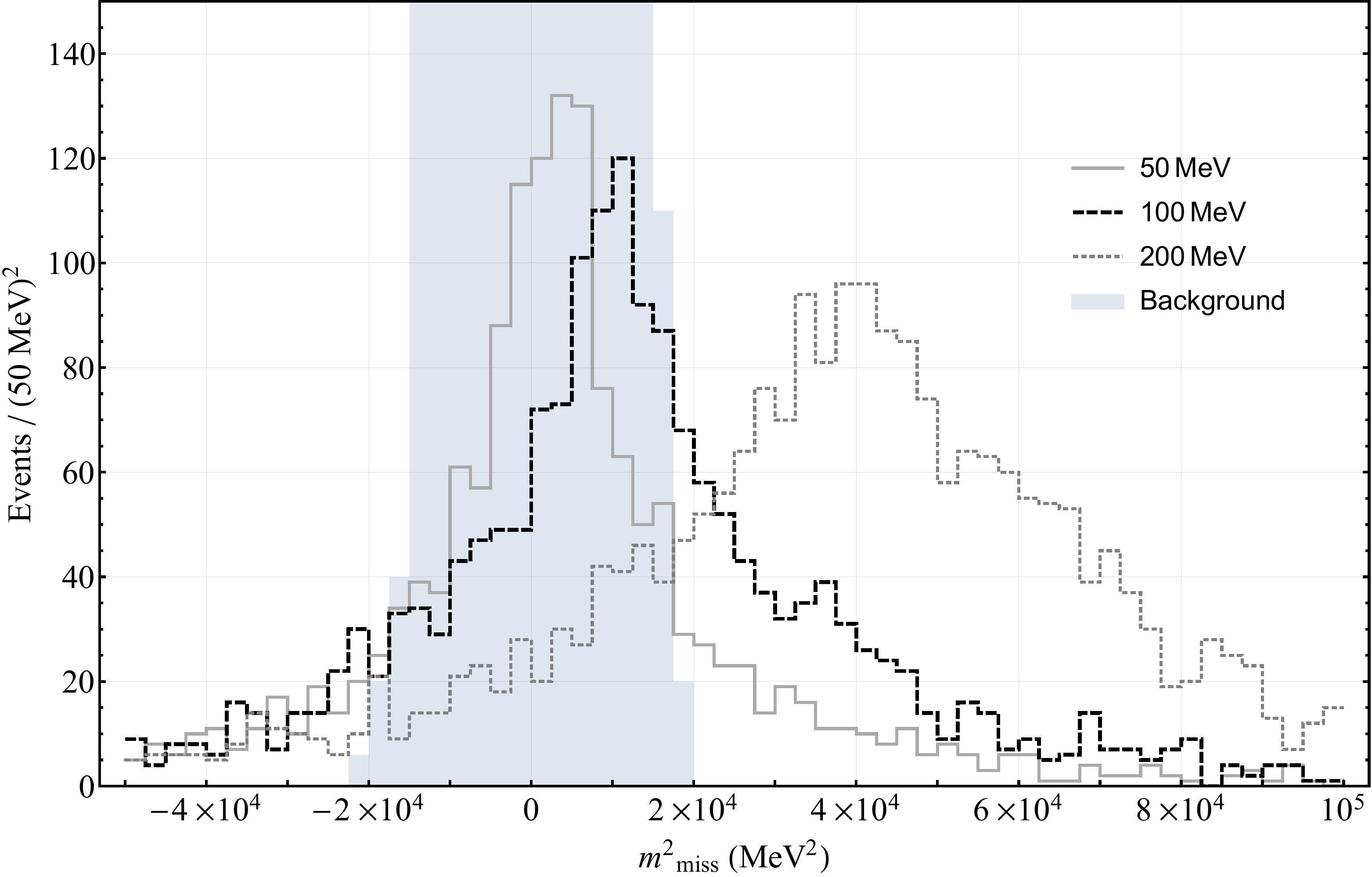}%
\vspace{-2mm}%
\caption{Distribution of missing mass square for both signal events $J/\psi \to
\mu^- \mu^+ X_1$ and background events $J/\psi \to \mu^- \mu^+
\gamma_\textrm{soft}$. (Here the background is composed of not only $J/\psi \to
\mu^- \mu^+ \gamma_\textrm{soft}$ but also experimentally smeared $J/\psi \to
\mu^- \mu^+$.) The central value of $g_1$ that solves $\Delta a_\mu$ for the
corresponding value of $m_X$ (central value of Fig.~\ref{fig:g01mX}) is used for
the demonstration. Shifts in $g_1$ will only scale the distribution accordingly.
The bin size of $50$~MeV for missing mass does not imply that mass of $X_1$ can
not be probed below $50$~MeV.}%
\label{fig:missing-mass-sq-vector}
\end{figure}
\begin{figure}[htb]
\centering%
\includegraphics[width=0.8\linewidth]{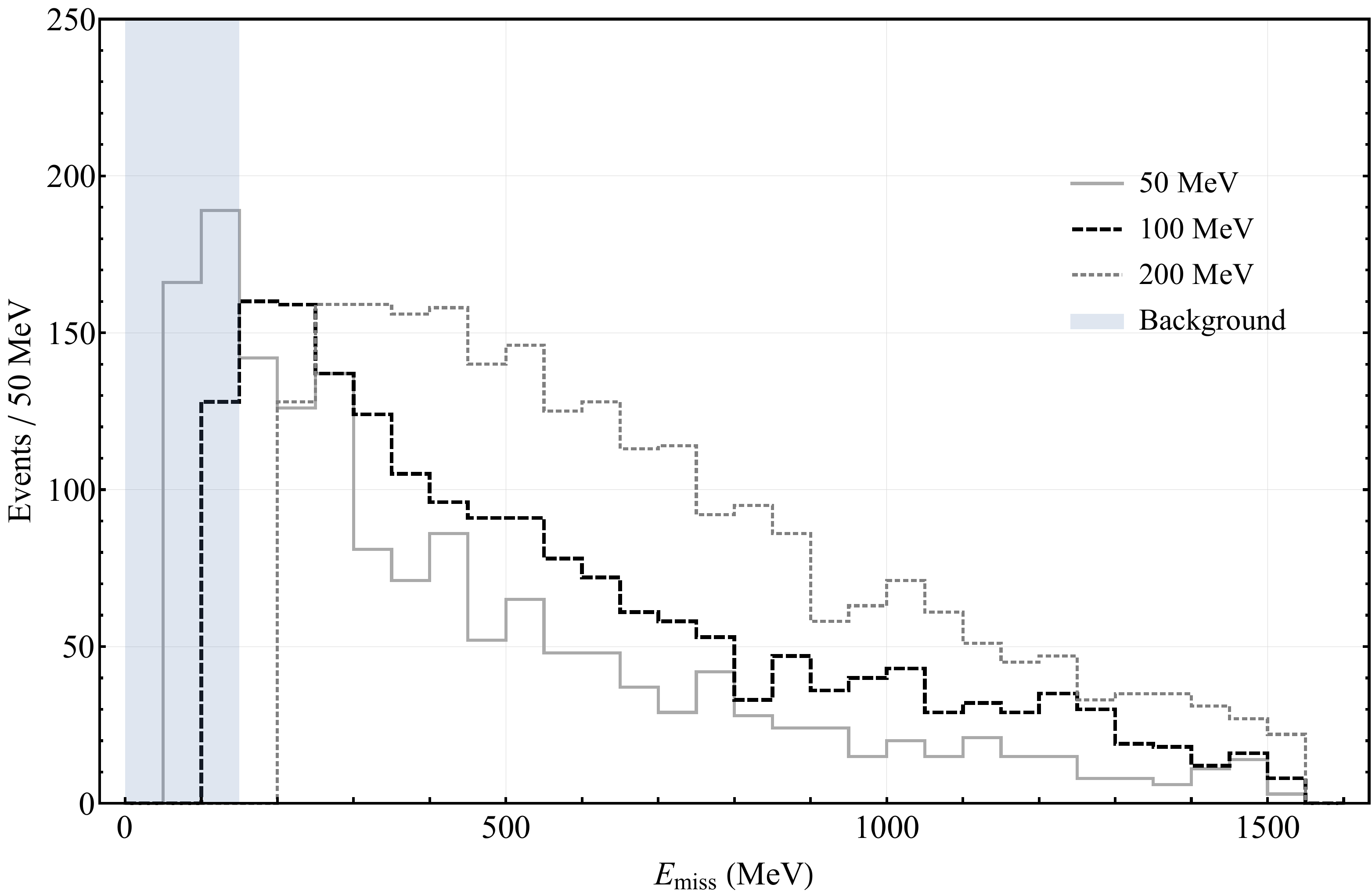}%
\vspace{-2mm}%
\caption{Distribution of missing energy for the signal events $J/\psi \to \mu^-
\mu^+ X_1$ and the background events $J/\psi \to \mu^- \mu^+
\gamma_\textrm{soft}$. Here the bin size for missing energy is $50$~MeV. All the
background events are confined to the region $E_\textrm{miss}<140$~MeV. Thus for
signal events with $m_X\lesssim 50$~MeV the missing energy peaks cannot be
observed after imposing the $E_{\textrm{miss}}< 140$~MeV cut. }%
\label{fig:bg-and-signal-energy-distribution}
\end{figure}

The results from our simulation for both signal and background events are shown
in Fig.~\ref{fig:missing-mass-sq-vector}. We have shown $m_\textrm{miss}^2$
distribution for three different signal cases corresponding to $m_X=50$~MeV,
$100$~MeV and $200$~MeV. It is easy to observe a corresponding shift in the
position of the signal peaks when we go from smaller mass to the larger mass.
The background is large in the neighborhood of $m_\textrm{miss}^2=0$, as
expected. It can be seen from Fig.~\ref{fig:missing-mass-sq-vector} that the
background events dominate up to around $m_{\textrm{miss}}^2\sim
(120~\textrm{MeV})^2$. This especially kills signal events with
$m_{\textrm{miss}}^2 < (120~\textrm{MeV})^2$ and makes it difficult to accurately
identify $m_X$, the mass of $X_1$. In order to identify $m_X$, we need to use
the missing energy information and apply energy cut as discussed below.

\subsubsection{Missing energy}
\label{subs:me}

The missing energy $E_{\textrm{miss}}$ is defined as
\begin{eqnarray} \label{eq:Emiss}
E_{\textrm{miss}} \equiv M_{J/\psi} - E_+ -E_-,
\end{eqnarray}
where $M_{J/\psi}$ is the mass of $J/\psi$, $E_\pm$ denote the energies of
$\mu^\pm$. The $E_{\textrm{miss}}$ distribution of a simulated background as well
as few benchmark signal cases (corresponding to $m_X=50$~MeV, $100$~MeV and
$200$~MeV) are shown in Fig.~\ref{fig:bg-and-signal-energy-distribution}. The
figure shows that the energy cut around $140$~MeV will completely eliminate the
background events (as the missing soft-photon would have energy $< 20$~MeV at
BESIII), while leaving most of signal events.

\begin{table}[tb]
\centering%
\begin{tabular}{cccc}\hline%
$m_X\textrm{[MeV]}$ & $\overline{\textrm{Br}}$ & $N_{\textrm{signal}}$ &
$N_{\textrm{signal}}^{\textrm{cut}}$ \\ [0.5ex] \hline%
50 & 0.032505 & 1546 & 1236 \\%
100 & 0.022285 & 1840 & 1747 \\%
150 & 0.016889 & 2127 & 2127 \\%
200 & 0.010947 & 2378 & 2378 \\ \hline%
\end{tabular}
\caption{Canonical branching fractions and the number of signal events before
and after applying the the missing energy cut, $E_{\textrm{miss}}=140$~MeV. The
central value of $g_1$ that solves $\Delta a_\mu$ for the corresponding value of
$m_X$ (central value of Fig.~\ref{fig:g01mX}) is used for computing the number
of signal events.} %
\label{table:1}
\end{table}

\begin{figure}[hbtp]
\centering%
\includegraphics[width=0.8\linewidth]{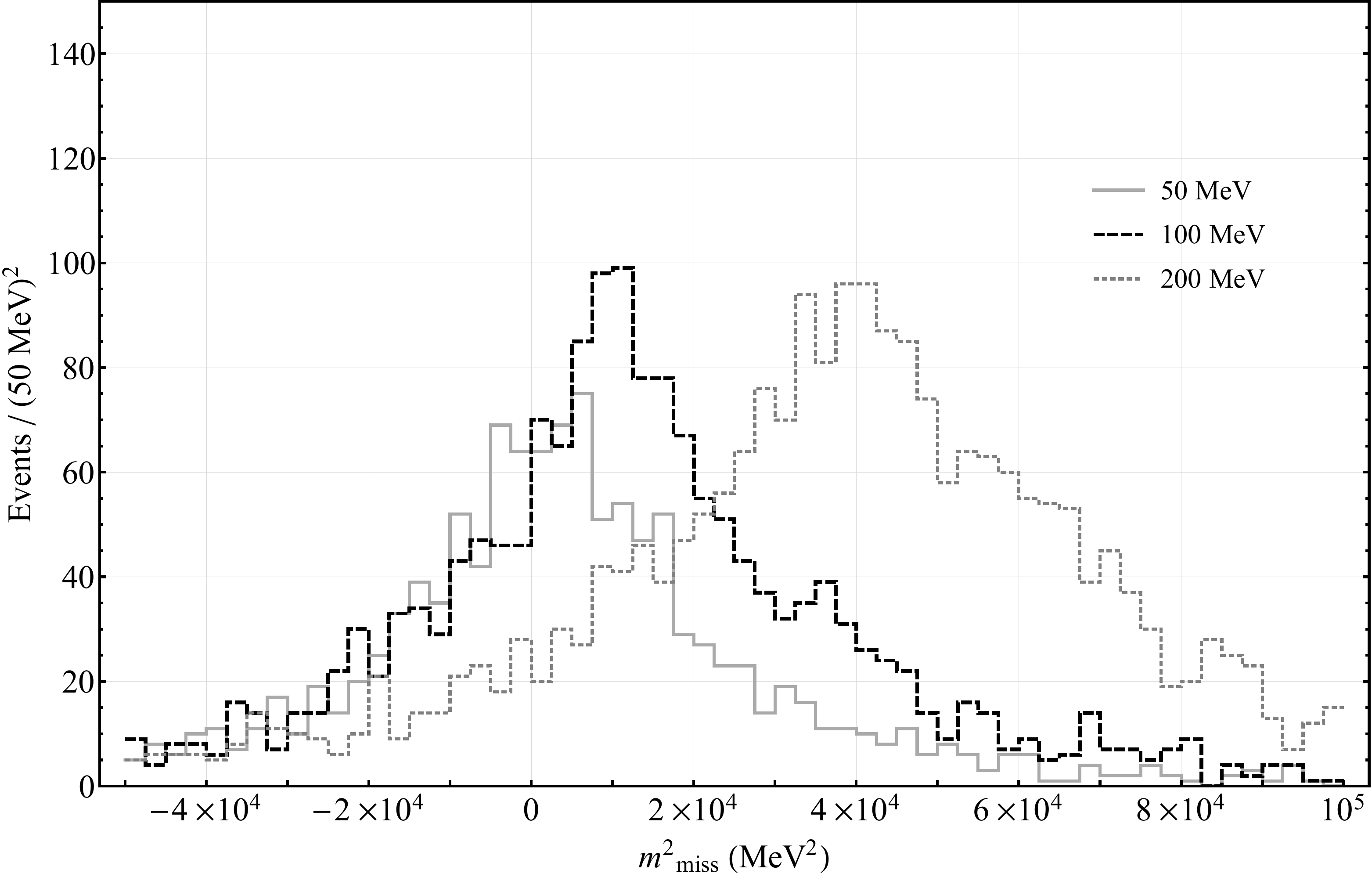}%
\vspace{-3mm}%
\caption{Distribution of missing mass square after applying
the minimum missing energy cut of $140$~MeV. }
\label{fig:m2-bg-signal-together-1011}
\end{figure}

In Table~\ref{table:1} we list the canonical branching ratio,
$\overline{\textrm{Br}}\equiv \textrm{Br}/g_1^2$, and the number of signal
events before and after applying the missing energy cut by which we throw away
all those events with $E_{\textrm{miss}}<140$~MeV. Only the number of signal
events with lighter $m_X$ (around $50$~MeV) is significantly reduced. After
applying the missing energy cut the $m_{\textrm{miss}}^2$ distributions gets
modified as is shown in Fig.~\ref{fig:m2-bg-signal-together-1011}. The figure
shows clearly that no background events survive after this cut is imposed. Now the new gauge boson mass can be extracted from the resultant distribution, shown in Fig.~\ref{fig:m2-bg-signal-together-1011}.
Please note that although the demonstrations were made with the specific values
of $g_1$, which solve the muon anomalous magnetic moment for corresponding
$m_X$, the strategy is generic and changes in $g_1$ will end up with nothing
more than overall scaling of the number of events (in
Fig.~\ref{fig:missing-mass-sq-vector},
\ref{fig:bg-and-signal-energy-distribution},
\ref{fig:m2-bg-signal-together-1011}, \ref{fig:delta-m-bg-signal-together-1011}
and Table~\ref{table:1}), while leaving the overall shapes of distributions
unchanged.

\subsubsection{Mass shift of muon pair from \texorpdfstring{$J/\psi$}{J/psi}}
\label{ssub:ms}

\begin{figure}[htb]
\centering%
\includegraphics[width=0.8\linewidth]{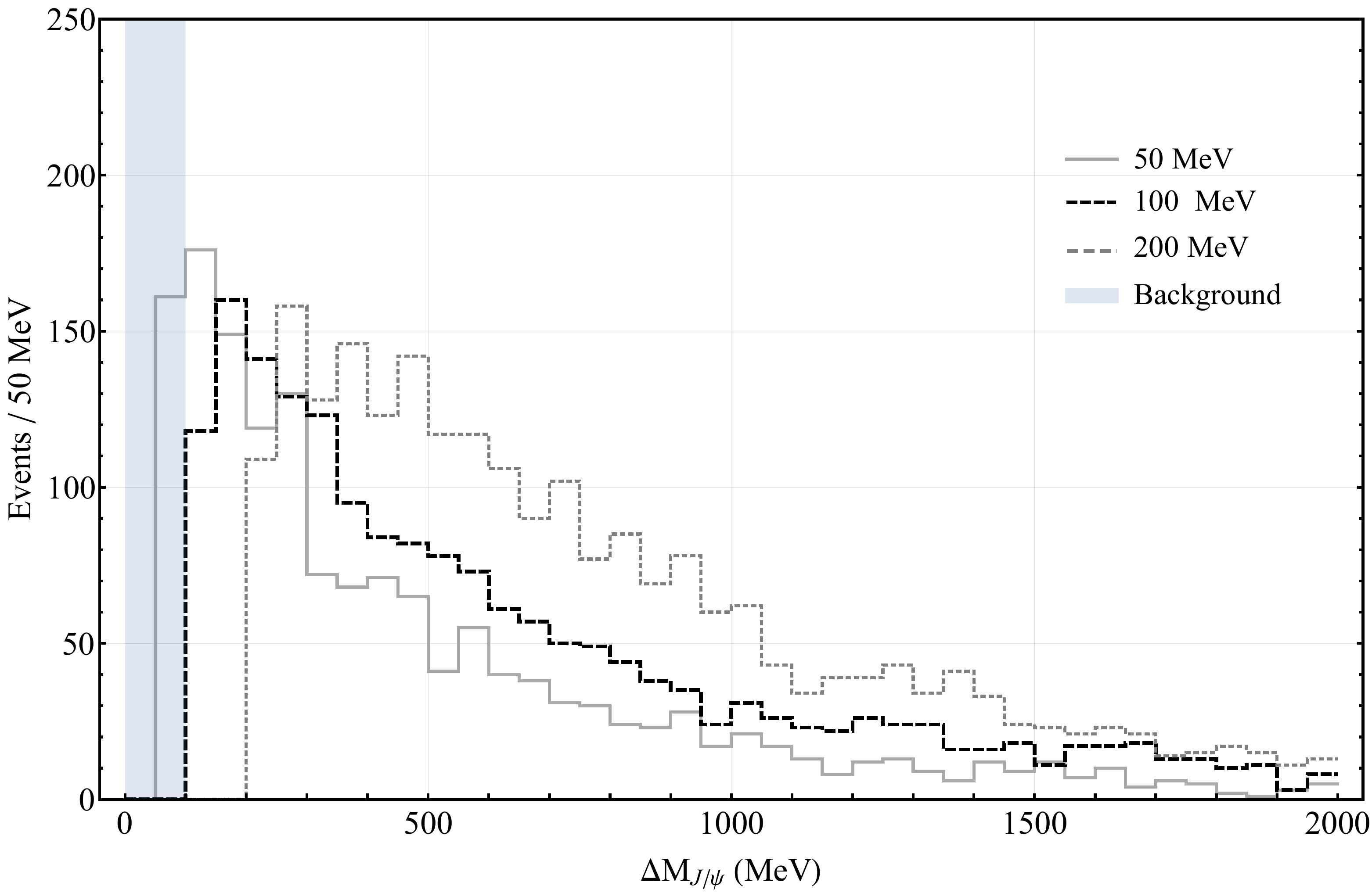}%
\vspace{-2mm}%
\caption{Distribution of the mass shift $\Delta(M_{J/\psi})$
for the signal events $J/\psi \to \mu^- \mu^+ X_1$ and the
background events $J/\psi \to \mu^- \mu^+ (\gamma_\textrm{soft})$. The effect of
background is negligible here, considering the long tail of the
$\Delta(M_{J/\psi})$ distribution of the signal. The background goes up to
around 70 MeV and it can be clearly distinguished from the signal for heavier
$X_1$ cases. }%
\label{fig:delta-m-bg-signal-together-1011}
\end{figure}

Alternatively one can utilize another observable to probe $X_1$, namely the
deviation in the measurement of invariant mass of observed muon pair from mass
of $J/\psi$,
\begin{eqnarray}\label{eq:DMSM}
\Delta(M_{J/\psi})\equiv M_{J/\psi}-\sqrt{(p_+ + p_-)^2}.
\end{eqnarray}
If this value significantly deviates from zero for a distribution of events,
those events would qualify as signal events. In the case of background event
$J/\psi \rightarrow \mu^- \mu^+ \gamma_\textrm{soft}$, $\Delta(M_{J/\psi})$
indicates difference between the actual and observed mass of $J/\psi$. So it
will be peaked at zero with some smearing due to error in momentum measurement.
Taking the energy and momentum resolution at BESIII experiment, the standard
deviation of $\Delta(M_{J/\psi})$ distribution comes around $11$~MeV and so
there will be some events up to around $70$~MeV. This is shown in the
Fig.~\ref{fig:delta-m-bg-signal-together-1011}.

It is easy to show that
\begin{equation}\label{eq:mass-shift}
\Delta(M_{J/\psi}) \simeq M_{J/\psi}-\sqrt{M_{J/\psi}^2+m_X^2-2M_{J/\psi}\,
E_X},
\end{equation}
where $E_X$ is the energy of $X_1$ and it is the same as $E_\textrm{miss}$, the
latter was analyzed in Sec.~\ref{subs:me} and
Fig.~\ref{fig:bg-and-signal-energy-distribution} for a few benchmark signal
scenarios. The minimum value of measured $\Delta(M_{J/\psi})$ is equal to $m_X$
when we substitute the minimum value of $E_X=m_X$ in Eq.~\eqref{eq:mass-shift}.
Thus the mass of $X_1$ can, in principle, be inferred from
Fig.~\ref{fig:delta-m-bg-signal-together-1011} by reading the minimum value of
$\Delta\left(M_{J/\psi}\right)$ for the corresponding distribution. However,
this information is also subject to the smearing effect from momentum
resolution. Nevertheless, looking at the $\Delta(M_{J/\psi})$ distribution would
complement our search for new physics using previously discussed observables. An
important feature of the $\Delta(M_{J/\psi})$ distribution is that it runs up to
much larger values for a signal compared to the background, even after smearing
is taken into account. Especially, if the mass of $X_1$ is larger than $70$~MeV,
the signal can be easily distinguished from the background.

\subsection{Probing the scalar case}

\begin{table}[ht!]
\centering%
\begin{tabular}{ccc}\hline%
$m_X\textrm{[MeV]}$ & $\overline{\textrm{Br}}_{\textrm{scalar}}$ & $\overline{\textrm{Br}}_{\textrm{vector}}$ \\[0.5ex] \hline%
50 & 0.0033254 & 0.032505 \\%
100 & 0.0022872 & 0.022285 \\%
150 & 0.0017978 & 0.016889 \\%
200 & 0.0014913 & 0.010947 \\%
\hline
\end{tabular}
\caption{Comparison of canonical branching ratios of $J/\psi \rightarrow \mu^-
\mu^+ X_{0,1}$ for a few chosen values of $m_X$.}%
\label{table:2}
\end{table}

\begin{figure}[ht]
\centering%
\includegraphics[width=0.99\linewidth,keepaspectratio]{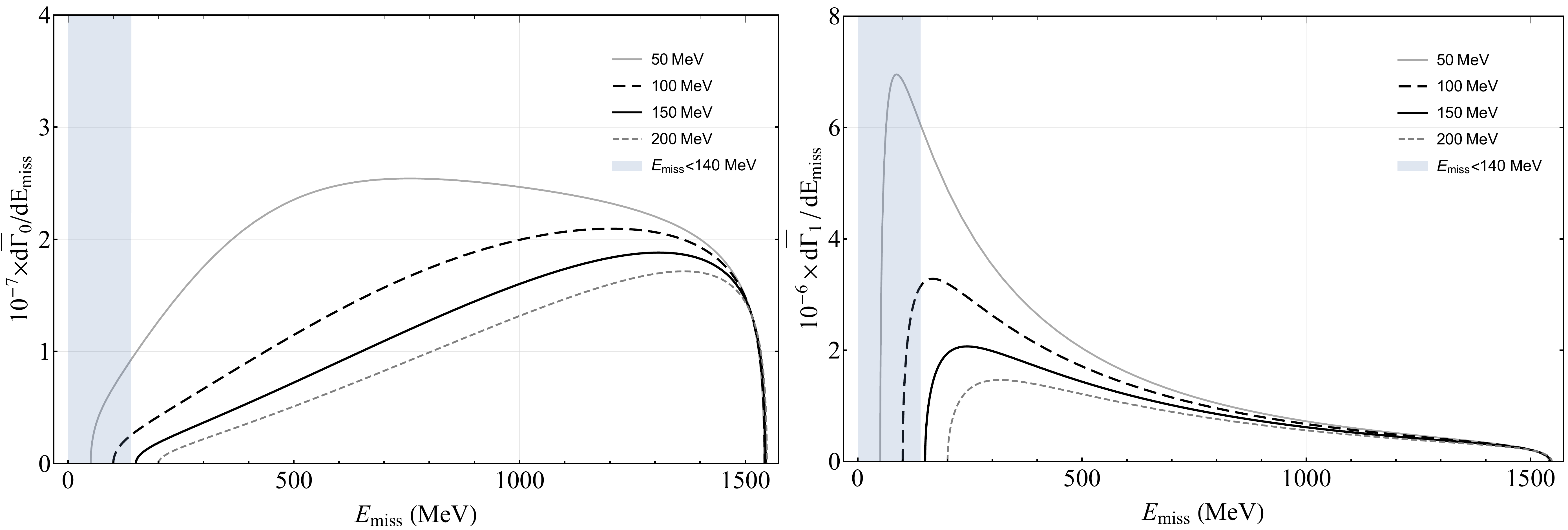}%
\vspace{-2mm}%
\caption{The unnormalized canonical missing energy distributions for scalar
$d\overline{\Gamma}_0/dE_{\textrm{miss}}$ and vector
$d\overline{\Gamma}_1/dE_{\textrm{miss}}$ new physics cases with different values
of $m_X$. The scalar cases exhibit a preference for higher missing energy as
opposed to the vector cases.} \label{fig:m2-bg-signal-together-1012}
\end{figure}

Comparing the canonical branching ratios of the scalar and vector cases, we find
that in the scalar case they are about 10 times smaller than in the vector case,
see Table~\ref{table:2}. Nevertheless, applying the same techniques as discussed
above for the vector case, we can also probe the scalar new physics possibility.
It is important to note that the missing energy distribution of a scalar is
different from that of a vector case. In
Fig.~\ref{fig:m2-bg-signal-together-1012} we plotted the canonical differential
decay rates $d\overline{\Gamma}_{0,1}/dE_\textrm{miss}$ (where
$\overline{\Gamma}_{0,1} \equiv \Gamma / g_{0,1}^2$ is the canonical decay width)
for the scalar and vector cases, respectively.  It is clear that scalar new
physics prefers higher missing energy whereas vector new physics prefers lower
missing energy. Thus, one can easily distinguish them in an experiment once
either of them gets detected.

\section{Conclusion}\label{sec:conclusion}

We investigated the possible $J/\psi \to \mu^+ \mu^- X$ events at the BESIII
experiment, where $X$ is a vector (or scalar) ``muon-philic'' particle which
could explain the present discrepancy $\Delta a_{\mu} \equiv a_{\mu}^{\rm exp} -
a_{\mu}^{\rm SM}$ of the anomalous magnetic moment of $\mu$ lepton,
cf.~Eq.~(\ref{eq:DelA}). It turns out that, if the coupling of $X$ to muon is
$g_{0,1} \sim 4 \times 10^{-4}$--$10^{-3}$ and its mass is  $m_X < 2 m_{\mu}$,
this particle can explain the mentioned discrepancy $\Delta a_{\mu}$.

The advantage of BESIII, in comparison with other experiments (Belle~II and
BaBar) where the continuum process $e^+ e^- \to \mu^+ \mu^- X$ is considered, is
that at BESIII there will be produced a very large number ($\sim 10^{11}$) of
on-shell $J/\psi$ particles (at rest) without any initial soft photon radiation.
Thus, the number of $J/\psi \to \mu^+ \mu^- X$ events can be significantly
higher than at Belle II or BaBar, because the latter experiments do not have
on-shell intermediaries for the continuum process $e^+ e^- \to \mu^+ \mu^- X$.
Further, the final state kinematics is more constrained at BESIII because of the
mentioned on-shellness and the very small decay width of $J/\psi$, making the
background effects smaller and easier to analyze than at Belle~II or BaBar. In
addition, the center-of-mass energy at Belle~II and BaBar is very high, which
makes the cross sections of $e^+ e^- \to \mu^+ \mu^- X$ suppressed ($\sigma
\propto 1/s$).

We showed that the number of events $J/\psi \to \mu^+ \mu^- X$ that take place
at BESIII in the mentioned range of parameters $(g_{0,1},m_X)$ is $\sim 10^3$
when $X$ is vector, and $\sim 10^2$ when $X$ is scalar. %
The main background to these events at BESIII is the final state radiation
$J/\psi \to \mu^+ \mu^- \gamma$. This is in contrast with Belle II
\cite{BelleII} (and BaBar) background to $e^+ e^- \to \mu^+ \mu^- X$, where
initial state radiation and $e^+ e^- \to \tau^+ \tau^- \to  \mu^+ \nu_{\mu}
{\bar \nu}_{\tau} \mu^-  {\bar \nu}_{\mu} \nu_{\tau}$ are additional background
sources. We showed that the distribution $m^2_{\rm miss}$, Eq.~(\ref{eq:mmiss}),
is a priori not a good quantity to identify the signal events $J/\psi \to \mu^+
\mu^- X$, because of strong background ($J/\psi \to \mu^+ \mu^- \gamma$)
contributions to this quantity. On the other hand, the distribution $d N_{J/\psi
\to \mu \mu X}/dE_{\rm miss}$, where $E_{\rm miss}=M_{J/\psi}- E_{+} - E_{-}$
[cf.~Eq.~(\ref{eq:Emiss})], is a good quantity to identify the signal events
when $m_X > 50$~MeV (and $m_X < 2 m_{\mu}$) once the cut $E_{\rm miss} <
140$~MeV is applied which eliminates the background; for $m_X < 50$~MeV, the
signal rate gets significantly diminished by the cut and the maximum is swamped
by the background. The mentioned cut $E_{\rm miss} < 140$~MeV also eliminates
completely the background to the quantity $m^2_{\rm miss}$. %
Further, we showed that the quantity $\Delta(M_{J/\psi})$, defined in
Eq.~(\ref{eq:DMSM}), is a good complementary quantity to identify the signal
events if $m_{X} > 70$~MeV. If $X$ is a scalar, the number of events in the
mentioned parameter range of $(g_{0,1},m_X)$ is by about a factor of 10 lower,
as mentioned earlier. However, the form of the distribution $d N_{J/\psi \to \mu
\mu X}/dE_{\rm miss}$ is in this case shifted to higher values of $E_{\rm
miss}$, allowing the scalar case to be easily distinguished from the vector
case.

In summary, in this paper we demonstrated that it is possible to not only probe
both the scalar and vector new physics cases contributing to anomalous magnetic
moment of muon by searching for the signal $J/\psi \to \mu^- \mu^+ +
\textrm{``missing,''}$ but also to distinguish between the SM background and the
new physics possibilities by using missing mass, missing energy and mass shift
of muon pair from $J/\psi$. Our numerical analysis clearly shows that BESIII
could be the best place, at present, to implement this study experimentally. The
amazing aspect of this probe is the possibility to either discover new physics
or to completely rule out the simplest explanations for the longstanding muon
anomalous magnetic moment discrepancy.

\acknowledgments The work of C.S.K., D.H.L. and D.S. was supported by the
National Research Foundation of Korea (NRF) grant funded by the Korean
government (MSIP) (NRF2018R1A4A1025334); the work of G.C. was supported by
FONDECYT (Chile) Grant No.~1180344.


\begin{thebibliography}{99}

\bibitem{Bennett:2006fi}%
G.~W.~Bennett {\it et al.} [Muon g-2 Collaboration], ``Final report of the muon
E821 anomalous magnetic moment measurement at BNL,''
Phys.\ Rev.\ D {\bf 73} (2006) 072003
[hep-ex/0602035].

\bibitem{Tanabashi:2018oca}%
M.~Tanabashi {\it et al.} [Particle Data Group], ``Review of Particle Physics,''
Phys.\ Rev.\ D {\bf 98} (2018) 030001.

\bibitem{Campanario:2019mjh}%
F.~Campanario, H.~Czy\.{z}, J.~Gluza, T.~Jeli\'{n}ski, G.~Rodrigo, S.~Tracz and
D.~Zhuridov, ``Standard model radiative corrections in the pion form factor
measurements do not explain the $a_\mu$ anomaly,'' Phys.\ Rev.\ D \textbf{100},
no.7, 076004 (2019) 
[arXiv:1903.10197 [hep-ph]].

\bibitem{Jegerlehner:2017lbd}%
F.~Jegerlehner, ``Muon $g-2$ theory: The hadronic part,'' EPJ Web Conf.\ 
\textbf{166} (2018), 00022 
[arXiv:1705.00263 [hep-ph]].

\bibitem{BelleII}%
Y.~Jho, Y.~Kwon, S.~C.~Park and P.~Y.~Tseng, ``Search for muon-philic new light
gauge boson at Belle II,'' JHEP {\bf 1910} (2019) 168
[arXiv:1904.13053 [hep-ph]].

\bibitem{BESIII}%
The BESIII collaboration have accumulated a sample of 10 billion $J/\psi$ events
together with a continuum data sample on 11 February 2019, see
\url{http://bes3.ihep.ac.cn/doc/3313.html}.

\bibitem{Yuan:2019zfo}%
C.~Z.~Yuan and S.~L.~Olsen, ``The BESIII physics programme,'' Nature Rev.\
Phys.\  {\bf 1}, no. 8, 480 (2019) 
[arXiv:2001.01164 [hep-ex]].

\bibitem{Ablikim:2019hff}%
M.~Ablikim {\it et al.} [BESIII Collaboration], ``Future Physics Programme of
BESIII,'' Chin.\ Phys.\ C \textbf{44}, no.4, 040001 (2020)
[arXiv:1912.05983 [hep-ex]].

\bibitem{Jiang:2018jqp}%
J.~Jiang, H.~Yang and C.~F.~Qiao, ``Exploring Bosonic Mediator of Interaction at
BESIII,'' Eur.\ Phys.\ J.\ C {\bf 79}, no. 5, 404 (2019)
[arXiv:1810.05790 [hep-ph]].

\bibitem{Correia:2016xcs}%
F.~Correia and S.~Fajfer, ``Restrained dark $U(1)_d$ at low energies,'' Phys.\
Rev.\ D \textbf{94}, no.11, 115023 (2016) 
[arXiv:1609.00860 [hep-ph]].

\bibitem{Correia:2019pnn}%
F.~Correia and S.~Fajfer, ``Light mediators in anomaly free $U(1)_{X}$ models.
Part I. Theoretical framework,'' JHEP \textbf{10}, 278 (2019)
[arXiv:1905.03867 [hep-ph]].

\bibitem{Correia:2019woz}%
F.~Correia and S.~Fajfer, ``Light mediators in anomaly free $U(1)_{X}$ models.
Part II. Constraints on dark gauge bosons,'' JHEP \textbf{10}, 279 (2019)
[arXiv:1905.03872 [hep-ph]].

\bibitem{Review}%
M.~Lindner, M.~Platscher and F.~S.~Queiroz, ``A call for New Physics : the muon
anomalous magnetic moment and lepton flavor violation,'' Phys.\ Rept.\  {\bf
731} (2018) 1 
[arXiv:1610.06587 [hep-ph]].

\bibitem{He:1990pn}%
X.~He, G.~C.~Joshi, H.~Lew and R.~Volkas, ``New $Z^\prime$ phenomenology,''
Phys.\ Rev.\ D \textbf{43} (1991) R22. 

\bibitem{He:1991qd}%
X.~He, G.~C.~Joshi, H.~Lew and R.~Volkas, ``Simplest $Z^\prime$ model,'' Phys.\
Rev.\ D \textbf{44}  (1991) 2118. 

\bibitem{Foot:1994vd}%
R.~Foot, X.~G.~He, H.~Lew and R.~R.~Volkas, ``Model for a light $Z^\prime$
boson,'' Phys.\ Rev.\ D {\bf 50} (1994) 4571 
[hep-ph/9401250].

\bibitem{Ruegg:2003ps}%
H.~Ruegg and M.~Ruiz-Altaba, ``The Stueckelberg field,'' Int.\ J.\ Mod.\ Phys.\
A {\bf 19} (2004) 3265 
[hep-th/0304245].

\bibitem{Feldman:2006wb}%
D.~Feldman, Z.~Liu and P.~Nath, ``The Stueckelberg $Z$ prime at the LHC:
discovery potential, signature spaces and model discrimination,'' JHEP {\bf
0611} (2006) 007 
[hep-ph/0606294].

\bibitem{Mishra:1991bv}%
S.~Mishra \textit{et al.} [CCFR], ``Neutrino tridents and $W$-$Z$
interference,'' Phys.\ Rev.\ Lett.\  {\bf 66}, 3117 (1991).

\bibitem{Altmannshofer:2014pba}%
W.~Altmannshofer, S.~Gori, M.~Pospelov and I.~Yavin, ``Neutrino trident
production: a powerful probe of new physics with neutrino beams,'' Phys.\ Rev.\
Lett.\  {\bf 113} (2014) 091801 
[arXiv:1406.2332 [hep-ph]].

\bibitem{kaonc} 
G.~Krnjaic, G.~Marques-Tavares, D.~Redigolo and K.~Tobioka, ``Probing muonic
forces and dark matter at kaon factories,'' Phys.\ Rev.\ Lett.\  {\bf 124}
(2020) no.4,  041802 
[arXiv:1902.07715 [hep-ph]].

\bibitem{Gauld:2013qja}%
R.~Gauld, F.~Goertz and U.~Haisch, ``An explicit Z'-boson explanation of the $B
\to K^* \mu^+ \mu^-$ anomaly,'' JHEP {\bf 1401}, 069 (2014)
[arXiv:1310.1082 [hep-ph]].

\bibitem{DAmico:2017mtc}%
G.~D'Amico, M.~Nardecchia, P.~Panci, F.~Sannino, A.~Strumia, R.~Torre and
A.~Urbano, ``Flavour anomalies after the $R_{K^*}$ measurement,'' JHEP {\bf
1709}, 010 (2017) 
[arXiv:1704.05438 [hep-ph]].

\bibitem{DAmbrosio:2019tph}%
G.~D'Ambrosio, A.~M.~Iyer, F.~Piccinini and A.~D.~Polosa, ``Confronting $B$
anomalies with low energy parity violation,'' Phys.\ Rev.\ D {\bf 101}, no. 3,
035025 (2020) 
[arXiv:1902.00893 [hep-ph]].

\bibitem{BBN1}%
A.~Kamada and H.~B.~Yu, ``Coherent propagation of PeV neutrinos and the dip in
the neutrino spectrum at IceCube,'' Phys.\ Rev.\ D {\bf 92} (2015) no.11, 
113004 
[arXiv:1504.00711 [hep-ph]].

\bibitem{BBN2}%
M.~Escudero, D.~Hooper, G.~Krnjaic and M.~Pierre, ``Cosmology with very light
$L_{\mu}-L_{\tau}$ gauge boson,'' JHEP {\bf 1903} (2019) 071
[arXiv:1901.02010 [hep-ph]].

\bibitem{Ablikim:2009aa}%
M.~Ablikim {\it et al.} [BESIII Collaboration], ``Design and Construction of the
BESIII Detector,'' Nucl.\ Instrum.\ Meth.\ A {\bf 614}, 345 (2010)
[arXiv:0911.4960 [physics.ins-det]].

\end{thebibliography}
\end{document}